\documentstyle[preprint,tighten,aps]{revtex}
\begin{document}
\draft
\title{
Algorithm for obtaining the gradient expansion of the
local density of states and the free energy of a superconductor
}

\author{Lorenz Bartosch and Peter Kopietz} 
\address{
Institut f\"{u}r Theoretische Physik, Universit\"{a}t G\"{o}ttingen,
Bunsenstrasse 9, D-37073 G\"{o}ttingen, Germany}
\date{February 2, 1999}
\maketitle
\begin{abstract}

We present an efficient algorithm
for obtaining the gauge-invariant gradient expansion of the local
density of states and the 
free energy of a clean superconductor.  
Our method is based on 
a new mapping of the semiclassical linearized Gorkov equations
onto a pseudo-Schr\"{o}dinger equation for a
three-component wave-function $\vec{\psi} (x )$, where
one component is directly related to the local density of states.
Because 
$\vec{\psi} ( x )$ satisfies a {\it{linear}} equation of motion,
successive terms in the gradient
expansion can be obtained by simple linear iteration.
Our method works equally well for real and complex
order parameter, and in the presence of
arbitrary external fields.
We confirm a recent
calculation of the fourth order
correction to the free energy by Kosztin, Kos, Stone and Leggett
[Phys. Rev. B {\bf{58}}, 9365 (1998)], 
who obtained a discrepancy with an earlier
result by Tewordt [Z. Phys. {\bf 180}, 385 (1964)].
We also give the fourth order correction to the local density of
states, which has not been published before.

\end{abstract}
\narrowtext
%
%
\section{Introduction}
%
The phenomenological Ginzburg-Landau theory
has proven to be a powerful method
to study the physical properties
of superconductors in spatially varying
external fields\cite{Tinkham96}.
The form of the
free energy $F \{ \Delta ( {\bf{r}} ) \}$
as a functional
of the complex superconducting order parameter
$\Delta ( {\bf{r}} )$ follows from general symmetry arguments.
Alternatively, 
for temperatures in the vicinity of
the critical temperature and for 
slowly varying external fields,
$F \{ \Delta ( {\bf{r}} ) \}$
can be derived microscopically from the Gorkov 
equations of superconductivity\cite{Abrikosov63}.
The microscopic approach can also be used to obtain corrections to the
Ginzburg-Landau free energy functional. 
Tewordt\cite{Tewordt64} explicitly calculated 
$F \{ \Delta ( {\bf{r}} ) \}$
up to fourth order in the gradients  of $\Delta ( {\bf{r}} )$.
By comparing the terms with four and two gradients,
he could estimate the
range of validity of the usual approximation
for the Ginzburg-Landau free energy
functional, where only terms with two 
gradients are
retained.
Unfortunately, a direct expansion of the free energy
of a superconductor in powers of gradients of $\Delta ( {\bf{r}} )$
is quite laborious, so that the result of Tewordt 
is rather difficult to verify, 
and the calculation of even higher orders seems almost
impossible.

Recently Kosztin, Kos, Stone and
Leggett \cite{Kosztin98}, and  
Kos and Stone \cite{Kos98} (KKSL) developed new and more 
efficient algorithms to obtain the gradient expansion
of the free energy $F \{ {\Delta} ( {\bf{r}} ) \}$.
Starting point of their calculations are the
Bogoliubov - de Gennes equations\cite{deGennes66} for the
two-component wave-functions and eigen-energies of
a superconductor in an external field. 
Assuming that $| \Delta ( {\bf{r}} )|$ is small compared with
the Fermi energy,
the low-energy physics can be obtained by linearizing
the energy dispersion in the
Bogoliubov - de Gennes equations. 
Then one arrives at the so-called
semi-classical Andreev equations\cite{Andreev64}, which
involve only first order derivatives.
In the work\cite{Kosztin98} KKSL  then use the special properties
of the Andreev equations to show that
for {\it{real}} $\Delta ( {\bf{r}} )$ the gradient expansion
of $F \{ \Delta ( {\bf{r}} ) \}$
can be indirectly obtained from the gradient
expansion of the resolvent of  
a one-dimensional Schr\"{o}dinger operator,
which is nothing but the 
square of the Andreev-Hamiltonian.
This resolvent satisfies a non-linear second order
differential equation, the so-called Gelfand-Dikii
equation\cite{Kosztin98,Kos98,Gelfand75}.
KKSL  proceed by solving the Gelfand-Dikii equation iteratively,
and then use the approximate solution to reconstruct the
gradient expansion of $F \{ \Delta ( {\bf{r}} ) \}$.
Although this strategy is more efficient than the direct
approach used by Tewordt\cite{Tewordt64}, 
this procedure still has the disadvantage
that, in order to obtain 
all contributions to $F \{ \Delta ( {\bf{r}} ) \}$
with $n$ gradients, one has to calculate
terms of higher order than $n$
in the iterative solution of the Gelfand-Dikii equation.
In the case of {\it{complex}} $\Delta ({\bf r})$ KKSL derive a $2
\times 2$ 
matrix generalization of the Gelfand-Dikii equation, which is
a form of the semiclassical Eilenberger 
equation\cite{Eilenberger64}. In this case no reshuffling of terms in
the iterative procedure is
necessary. To take care of a zero-mode KKSL only have to consider the
next order in the gradient expansion. While in the work
\cite{Kosztin98} gradients of the magnitude and the phase of $\Delta
({\bf r})$ are handled separately, an expansion in terms of 
gradients of  $\Delta ({\bf r})$ itself is given in \cite{Kos98}.

In this work we shall develop an alternative algorithm
for calculating the gradient expansion of the free energy
of a superconductor. 
We derive a pseudo-Schr\"{o}dinger equation which is equivalent to the
Eilenberger equation and develop a gradient expansion of the local
density of states. In this expansion there appears a zero-mode which
needs to be taken care of. In contrast to KKSL we fix this problem by
making use of a non-linear constraint which naturally appears in our
formalism. We do believe that this is an interesting and efficient
alternative to the methods developed by KKSL.
Our method works equally well
for real and complex $\Delta ( {\bf{r}} )$ and 
directly generates the gradient expansion of the local
density of states of the superconductor, from which
the free energy can be obtained by simple integration.
As an application of our method, we have explicitly calculated
the local density of states
$\rho ( {\bf{r}} , \omega )$ and the free energy
$F \{ \Delta ( {\bf{r}} ) \}$ up to fourth order in the
gradients.
We confirm the result by KKSL\cite{Kosztin98,Kos98} for the fourth
order correction to $F \{ \Delta ( {\bf{r}} ) \}$,
who obtained a discrepancy with the
expression published by Tewordt\cite{Tewordt64}.

\section{From the Gorkov equations 
to a pseudo-Schr\"odinger equation}

In this section we shall  map the 
problem of calculating the local density of states (DOS)
of a superconductor onto the problem of
calculating the {\it{time-evolution of the wave-function 
of an effective $J=1$ quantum spin 
in a time-dependent complex magnetic field}}.
For real $\Delta$ we have recently used 
a similar mapping to study the effect of order
parameter fluctuations in disordered Peierls chains\cite{Bartosch99}.
Here we generalize this mapping to the case of complex $\Delta$.

\subsection{Reduction to an effective one-dimensional problem}

Starting point of our manipulations are the Gorkov equations for the
matrix Green's function of a clean superconductor
in the magnetic field ${\bf H} ({\bf r}) = {\bf
  \nabla}_{\bf{r}} \times {\bf A}({\bf r})$, 
 \begin{equation}
 \left( 
 \begin{array}{cc}
 \omega - {\cal H}_{\bf r} &
 \Delta({\bf r}) \\ 
 \Delta^{\ast}({\bf r}) & \omega - {\cal H}^{\ast}_{\bf r}
 \end{array}
 \right ) {\cal{G}} ({\bf r} ,{\bf r}^{\prime} , \omega )
 =
 \delta ( {\bf r} - {\bf r}^{\prime} ) \sigma_0
 \; ,
 \label{eq:Gorkov}
 \end{equation}
where $\sigma_0$ is the $2 \times 2$ unit matrix, and
the differential operator ${\cal{H}}_{\bf r}$ is given by
 \begin{equation}
 {\cal H}_{\bf r} =  
 \frac{[-i{\bf \nabla}_{\bf{r}} +\frac{e}{c}{\bf A}({\bf r})]^2}{2m}
 - \mu  \; . 
 \end{equation}
We have set $\hbar = 1$ and measure
energies and frequencies with respect to the
chemical potential $\mu$. Here $c$ is the speed of light and $-e$ is the
charge of an electron.
Note that KKSL\cite{Kosztin98,Kos98} start from
the Bogoliubov - de Gennes equations (which are
eigenvalue equations for the {\it wave-functions} 
of the superconductor)
while our starting point are the Gorkov equations
(which are differential equations for the {\it{Green's functions}}).

If
$|\Delta({\bf r}) | \ll \mu$  and
if we are only interested in low-energy, long wavelength
properties of the superconductor, 
we may linearize the energy dispersion.
For calculating the local DOS and the free energy, we only need to know
the Green's function at coinciding points
${\bf{r}} = {\bf{r}}^{\prime}$.
Following Waxman\cite{Waxman93}, we write
\begin{equation}
 {\cal G} ({\bf r},{\bf r} , \omega) \approx
 \frac{\nu_3 }{ \nu_1 } 
 \left< {\cal G}_{\bf n} ({\bf r},{\bf r}, \omega) \right>_{\bf
   n}
   \; .
 \label{eq:threetoone}
\end{equation}
Here $\nu_d$ is the $d$-dimensional DOS of 
free fermions at the Fermi energy (including both spin
directions), and $\left< \ldots 
\right>_{\bf n} \equiv \int \ldots
\frac{d\Omega_{\bf n}}{4\pi} $ 
denotes directional averaging
over the directions of the three-dimensional unit vector
${\bf{n}}$. The auxiliary Green's 
function ${\cal G}_{\bf n} ({\bf r},{\bf r}^{\prime} , \omega)$ 
satisfies 
 \begin{eqnarray}
  \left( 
 \begin{array}{cc}
 \omega - V ( {\bf r} ) + i v_F {\bf n \cdot \nabla}_{\bf{r}} & 
 \Delta ( {\bf 
   r} ) \\ 
 \Delta^{\ast} ( {\bf r} ) &  \omega - V ( {\bf r} ) - i v_F {\bf n
   \cdot \nabla}_{\bf{r}}
 \end{array}
 \right) 
 & & 
 \nonumber \\  
 & & \hspace{-50mm}  \times
 {\cal{G}}_{\bf n} ( {\bf r} , {\bf r}^{\prime} , \omega )
 = \delta ( {\bf{n}} \cdot ( {\bf r} -  
 {\bf r}^{\prime} ) )  \sigma_0
 \label{eq:linearizedGorkov}
 \; ,
 \end{eqnarray}
where $V({\bf r}) = \frac{e}{c} {\bf n \cdot A} ({\bf r})$ and $v_F$
is the Fermi velocity.
Note that Eq.(\ref{eq:linearizedGorkov})
is the Green's function analog of the 
Andreev equations for the wave-function \cite{Andreev64}.
Introducing ${\bf r} = x {\bf n} + {\bf r}_{\perp}$ 
and $\partial_x \equiv {\bf n
  \cdot \nabla}_{\bf{r}}$  (where ${\bf{n}} \cdot {\bf{r}}_{\bot} = 0$),
dropping the dependence on the parameters ${\bf r}_{\perp}$ and ${\bf
  n}$, and setting 
for simplicity $v_F = 1$,
we can write Eq.(\ref{eq:linearizedGorkov})
as a truly
one-dimensional equation,
 \begin{eqnarray}
  \left( 
 \begin{array}{cc}
 \omega - V ( x ) + i  \partial_x & \Delta ( x ) \\
 \Delta^{\ast} ( x ) & \omega - V ( x ) - i  \partial_x 
 \end{array}
 \right )
 {\cal{G}} ( x , x^{\prime} , \omega ) \nonumber 
 \\ 
 & & \hspace{-20mm} = \delta ( x - x^{\prime} ) \sigma_0
 \label{eq:Andreev}
 \; .
 \end{eqnarray}
To eliminate the forward scattering
potential $V ( x )$ in Eq.(\ref{eq:Andreev}) we set
 \begin{equation}
 {\cal{G}} ( x , x^{\prime} , \omega )  =
 e^{- \frac{i}{2} \theta ( x ) \sigma_3 }
 {\cal{G}}_1 ( x , x^{\prime} , \omega )  
 e^{ \frac{i}{2} \theta ( x^{\prime} ) \sigma_3 }
 \label{eq:Gbardef}
 \; ,
 \end{equation}
where $\sigma_3$ is the usual Pauli matrix.
Choosing the real function $\theta ( x )$ such that
 \begin{equation}
 \frac{1}{2} \partial_x \theta (x) = V ( x )
 \; ,
 \end{equation}
we obtain
 \begin{equation}
 \left(
 \begin{array}{cc}
 \omega + i  \partial_x & \tilde{\Delta} ( x ) \\
 \tilde{\Delta}^{\ast} ( x)   & \omega - i  \partial_x 
 \end{array}
 \right )
 {\cal{G}}_1 ( x , x^{\prime} , \omega ) = \delta ( x - x^{\prime} )
 \sigma_0
 \label{eq:Andreev1}
 \; ,
 \end{equation}
with
 \begin{equation}
 \tilde{\Delta} ( x ) = \Delta ( x ) e^{i \theta ( x ) }
 \; .
 \label{eq:deltatildedef}
 \end{equation}

\subsection{Pseudo-Schr\"{o}dinger equation for the local DOS}

The gradient expansion of the free energy can be obtained
from the local DOS. 
Because
$ {\cal{G}}_1 ( x , x, \omega )$ has the same diagonal elements
as
$ {\cal{G}} ( x , x, \omega )$, 
the local DOS $\rho_1 ( x , \omega )$ of
our effective one-dimensional model can be written as
 \begin{equation}
 \rho_1 ( x , \omega ) = 
 - \pi^{-1}
 {\rm Im} {\rm Tr} [ 
 {\cal{G}}_1 ( x , x , \omega + i0 ) ]
 \; ,
 \label{eq:localdos}
 \end{equation}
where due to the trace we get a factor of $2$ which takes both spin
directions into 
account. Note that in the matrix equation (\ref{eq:Andreev1})
the derivative operator $\partial_x$ is
proportional to the Pauli matrix $\sigma_3$. 
In the following it will be advantageous if the derivative operator
$\partial_x$ is proportional to the unit matrix.
Using $\sigma_3^2 = \sigma_0$,  
we obtain from Eq.(\ref{eq:Andreev1})
a differential equation 
with this property by
setting\cite{Bartosch99,Kos98},
 \begin{equation}
 {\cal{G}}_2 ( x , x^{\prime} , \omega ) 
 = \sigma_3
 {{\cal{G}}}_1 ( x , x^{\prime} , \omega ) 
 \; ,
 \label{eq:tildeGdef}
 \end{equation}
so that Eq.(\ref{eq:Andreev1}) implies
 \begin{eqnarray}
\left[ i  \partial_x \sigma_0 + \omega \sigma_3 -
  \tilde{\Delta} (x) \sigma_{+} + \tilde{\Delta}^{\ast} (x)
      \sigma_{-} \right]  {{\cal{G}}}_2 ( x , x^{\prime}  )   
 & & 
 \nonumber 
 \\
 & & \hspace{-20mm}
 = \delta ( x - x^{\prime} ) \sigma_0
 \label{eq:matrix}
 \; .
 \label{eq:Glin}
 \end{eqnarray}
For simplicity we have omitted the frequency label. 
The three Pauli matrices are denoted by
$\sigma_{i}$, $i= 1,2,3$, and  $\sigma_{\pm} =
\frac{1}{2 } [ \sigma_{1} \pm i \sigma_{2} ]$.  
We now generalize the method described 
for real $\Delta$
in Ref.\cite{Bartosch99}
to the case of complex $\Delta$.
We start by making the {\it{non-Abelian 
Schwinger-ansatz}}\cite{Bartosch99,Schwinger62},
 \begin{equation}
 {{\cal{G}}}_2 ( x , x^{\prime}  ) = U ( x )
 {\cal{G}}_0 ( x , x^{\prime}  )  U^{-1} ( x^{\prime} )
 \; ,
 \label{eq:ansatz}
 \end{equation}
where $U ( x )$ is an invertible $2 \times 2$ matrix.
It is easy to see that the solution of Eq.(\ref{eq:Glin})
can indeed be written in this form provided
${\cal{G}}_0$ and $U$ satisfy
 \begin{equation}
 \left[ i \partial_x \sigma_0 +\omega  \sigma_{3}
    \right]
 {\cal{G}}_0 ( x , x^{\prime}  )
  =  \delta ( x - x^{\prime}) \sigma_0
 \label{eq:matrix1}
 \; ,
 \end{equation}
 \begin{eqnarray}
 i \partial_x U  (x)  & = & \omega
 [ U (x) \sigma_3 - \sigma_3 U ( x ) ] 
 \nonumber
 \\
 & & 
 + 
 [ \tilde{\Delta} (x) \sigma_{+} - 
 \tilde{\Delta}^{\ast} (x) \sigma_{-} ]   U ( x )
 \; .
 \label{eq:Udif}
 \end{eqnarray}
We parameterize $U (x)$ as follows\cite{Bartosch99,Schopohl98},
 \begin{equation}
 U ( x ) = 
 e^{i \Phi_{+} ( x ) \sigma_{-} }
 e^{i \Phi_{-} ( x ) \sigma_{+} }
 e^{i \Phi_{3} ( x ) \sigma_{3} }
 \; .
 \label{eq:Euler}
 \end{equation}
This leads to the following system of equations
for the complex functions
$\Phi_{\pm} ( x )$ and
$\Phi_{3} ( x )$,
\begin{mathletters}
 \begin{eqnarray}
 \partial_x \Phi_{+} & = & - 2 i \omega \Phi_{+} 
 + \tilde{\Delta}^{\ast} - \tilde{\Delta} \Phi_{+}^2  
 \label{eq:phiplus}
 \; ,
 \\
 \partial_x \Phi_{-} & = &  2 i \omega \Phi_{-} - \tilde{\Delta}
 [ 1 - 2 \Phi_{+} \Phi_{-} ] 
 \label{eq:phiminus}
 \; ,
 \\
 \partial_x \Phi_{3} & = & - i \tilde{\Delta} \Phi_{+}
 \label{eq:phi3}
 \; .
 \end{eqnarray}
\end{mathletters}
For real $\tilde{\Delta}$
these equations
reduce to the set of equations given in Ref.\cite{Bartosch99}.
Recently Schopohl\cite{Schopohl98}
used an analogous procedure to map the semiclassical
Eilenberger equations\cite{Eilenberger64}
onto a similar set of non-linear equations.

The one-dimensional local DOS defined in Eq.(\ref{eq:localdos})
can now be written as
 \begin{equation}
 \rho_1 ( x , \omega ) = \nu_{1} {\rm Re} R ( x , \omega + i 0 )
 \; ,  
 \label{eq:rhoR}
 \end{equation}
where the complex variable $R ( x )$ is defined by
(suppressing again the frequency label)
 \begin{equation}
 R ( x  )  = 1 - 2 \Phi_{+} ( x  ) \Phi_{-}  ( x )
 \; .
 \label{eq:Rdef}
 \end{equation}
In principle one could now try to solve 
Eqs.(\ref{eq:phiplus}) and (\ref{eq:phiminus})
iteratively in powers of the derivatives of $\tilde{\Delta} ( x )$,
and then obtain the gradient expansion of the local DOS
from Eqs.(\ref{eq:rhoR}) and (\ref{eq:Rdef}).
However, the structure of the iterative
solution becomes more transparent if we introduce
the three-component complex vector 
 \begin{equation}
  \vec{\psi} ( x ) =
 \left(
 \begin{array}{c}
 Z_{+} ( x )  \\ R (x ) \\ Z_{-}  ( x )
 \end{array} 
 \right)
 =  
 \left(
   \begin{array}{c}
     - \sqrt{2} [ 1 - \Phi_{+} (x) \Phi_{-} (x) ] \Phi_{+} (x) \\
     1 - 2 \Phi_{+} ( x  ) \Phi_{-}  ( x ) \\
     \sqrt{2}  \Phi_{-} (x)
   \end{array}  
 \right) 
 \; ,
 \label{eq:psidef}
 \end{equation}
such that the $2 \times 2$ matrix Green's function at coinciding space
points,
\begin{equation}
  \label{govx}
  g(x) \equiv \frac{1}{2}[{\cal{G}}_{2}(x+0^{+},x)+{\cal{G}}_{2}(x,x+0^{+})]
\end{equation}
is equal to
\begin{equation}
  \label{eq:g}
  g(x)=\frac{i}{2}R(x)\sigma_{3}+\frac{1}{\sqrt{2}}[Z_{-}(x)\sigma_{+}+Z_{+}(x)\sigma_{-}] \; . 
\end{equation}
Because the components of $\vec{\psi}$
can be parameterized by only two variables
$\Phi_{\pm}$, they satisfy a constraint.
Introducing the row vector
 \begin{equation}
 \tilde{\psi}^{T} ( x ) = [ - Z_{-} ( x ) , R ( x ) , - Z_{+} ( x ) ]
 \; ,
 \label{eq:psitildedef}
 \end{equation}
the constraint can be simply written as
 \begin{equation}
 \tilde{\psi}^{T} ( x ) \vec{\psi} ( x ) = 
 R^2 ( x )  -  2 Z_{+} (x) Z_{-} (x ) = 1 \; ,
 \label{eq:constraint2}
 \; 
 \end{equation}
which can also be expressed as
\begin{equation}
 \label{eq:constraint3}
  g^2(x) = -\frac{1}{4}\sigma_{0} \; .
\end{equation}
Differentiating the vector $\vec{\psi}(x)$ we find the
equation of motion
 \begin{equation}
  - \partial_x \vec{\psi}  ( x ) = H ( x )  \vec{\psi} ( x )
  \; ,
  \label{eq:pseudoschroedinger}
  \end{equation}
with the pseudo-Hamiltonian given by
 \begin{equation}
 H (x )  = 
 \left(
 \begin{array}{ccc}
   2 i \omega    & \sqrt{2}  \tilde{\Delta}^{\ast} (x) & 0  \\
  \sqrt{2} \tilde{\Delta} ( x )  &  0 & \sqrt{2} \tilde{\Delta}^{\ast} ( x )  \\
   0 &  \sqrt{2} \tilde{\Delta} ( x )  &  - 2 i \omega 
  \end{array}
  \right)
  \; .
 \label{eq:multistoch}
 \end{equation}
Note that $H ( x )$ can also be written as
 \begin{equation}
  H (x ) 
   =  
 2 i  \omega J_3 +  \tilde{\Delta} ( x )  J_{-}
 + \tilde{\Delta}^{\ast}  ( x ) J_{+}
 \; , 
 \label{eq:Hdef}
 \end{equation}
where the $J_i$ are spin $J=1$ operators in the representation
 \begin{equation}
 J_3  = 
 \left(
 \begin{array}{ccc}
   1   & 0 & 0 \\
  0 &  0 &  0 \\
   0 &  0 &  -1 
  \end{array}
  \right)
  \; \; ,  \; \; 
  \\
 J_{+}  
 = {\sqrt{2}}
 \left(
 \begin{array}{ccc}
   0   & 1 & 0 \\
  0 &  0 &  1 \\
   0 & 0  &  0 
  \end{array}
  \right)
  =
 J_{-}^{\dagger} 
  \; . 
  \label{eq:Ldef}
  \end{equation}
Using the alternative parameterization of $\vec{\psi}(x)$
in terms of the $2 \times 2$ matix $g(x)$ given in (\ref{eq:g}), we
find that our 
pseudo-Schr\"{o}dinger equation is equivalent to the Eilenberger equation,
\begin{equation}
  \label{eq:Eilenberger}
  \partial_{x} g(x) = [i\omega \sigma_{3}-i\tilde{\Delta}(x)
  \sigma_{+} + i\tilde{\Delta}^{\ast}(x) \sigma_{-},g(x)] \; .
\end{equation}
This can be seen by using the commutation relations 
for the Pauli matrices and reducing Eq. (\ref{eq:Eilenberger}) to the
equation of 
motion (\ref{eq:pseudoschroedinger}). Thus we have found an
unconventional derivation of the Eilenberger equations which is based
on the non-Abelian Schwinger-ansatz.

Formally Eq.(\ref{eq:pseudoschroedinger}) looks
like the imaginary time Schr\"{o}dinger equation
for a $J=1$ quantum spin subject to an imaginary
time-dependent magnetic field. 
Recall that the real part of the second component of the
state $\vec{\psi} ( x )$ can be identified with the
local DOS. Because our
pseudo-Schr\"{o}dinger equation is linear,
the gradient expansion of the local DOS
can now be generated by a straightforward 
iterative calculation of the 
state $\vec{\psi} ( x )$ in powers of gradients.
Let us emphasize that, apart from the semiclassical approximation
(Eq.(\ref{eq:threetoone})), so far no approximation
has been made. We have simply mapped
the original problem onto an effective
pseudo-Schr\"{o}dinger equation, which is
the most convenient starting point for
setting up the gradient expansion.

\section{Recursive algorithm and gradient expansion}

The gradient expansion of the local DOS
is directly obtained from the second component of the gradient
expansion of
$\vec{\psi} ( x )$.
For convenience we
develop the gradient expansion of $\vec{\psi } ( x )$ 
for imaginary frequencies $\omega = i E$, because then
our pseudo-Hamiltonian (\ref{eq:multistoch}) is Hermitian and left and
right 
eigenvectors are identical. 
Suppose we expand the solution of Eq.(\ref{eq:pseudoschroedinger})
in the form
 \begin{equation}
 \vec{\psi} ( x  ) = \sum_{n = 0}^{\infty} \vec{\psi}_n ( x )
 \; ,
 \label{eq:psiexp}
 \end{equation}
where by definition  $\vec{\psi}_n ( x )$ involves $n$ derivatives with 
respect to $x$. Obviously 
 \begin{equation}
 H ( x ) \vec{\psi}_0 ( x ) = 0
 \; ,
 \label{eq:psi0}
 \end{equation}
i.e. 
$\vec{\psi}_0 ( x )$ must be an eigenvector
of $H (x)$ with eigenvalue zero. The existence of such an
eigenvector follows trivially from the fact
that our pseudo-Hamiltonian  (\ref{eq:Hdef})
can be interpreted as the
Zeeman-Hamiltonian of a $J =1$ quantum spin in an external magnetic
field. 
Note that Eq.(\ref{eq:psi0}) determines 
$\vec{\psi}_0 ( x )$ only up to an overall multiplicative factor, which is
fixed by requiring that the components
of $\vec{\psi}_0 ( x )$ satisfy 
the constraint (\ref{eq:constraint2}).
This yields (with $\omega = i E $)
 \begin{equation}
 \vec{\psi}_0 ( x ) = \frac{1}{\sqrt{ E^2 + | \tilde{\Delta} ( x ) |^2 }}
 \left( \begin{array}{c} 
 \frac{\tilde{\Delta}^{\ast} ( x )}{\sqrt{2}}
 \\ E \\ 
 - \frac{\tilde{\Delta} ( x )}{\sqrt{2}}
 \end{array}
 \right)
 \; .
 \label{eq:psi02}
 \end{equation}
For the higher order terms we obtain the simple recursion relation
 \begin{equation}
 \partial_x \vec{\psi}_n ( x ) = - H ( x ) \vec{\psi}_{n+1} ( x )
 \; \; , \; \; n = 0,1, \ldots
 \; .
 \label{eq:recursion}
 \end{equation}
Because one of the eigenvalues of $H ( x )$ vanishes,
the inverse of $H ( x ) $ does not exist, so that we cannot simply solve
Eq.(\ref{eq:recursion}) by multiplying both sides by
$H^{-1} ( x )$. As a consequence, Eq.(\ref{eq:recursion})
determines $\vec{\psi}_{n+1} ( x )$ only up to 
a vector proportional to $\vec{\psi}_0 ( x )$,
 \begin{equation}
 \vec{\psi}_{n+1} ( x )  =  - H_{\bot}^{-1} ( x ) \partial_x
 \vec{\psi}_n ( x ) 
 + c_{n+1} ( x ) \vec{\psi}_0 ( x )
 \; ,
 \label{eq:psinplus1}
 \end{equation}
where $H_{\bot}^{-1} ( x )$ is the inverse
of $H (  x) $ in the subspace orthogonal to
$\vec{\psi}ß_0 ( x)$. Using the fact that the two
non-vanishing eigenvalues of $H(x)$ are
given by $\pm 2 [ E^2 + | \tilde{\Delta} ( x ) |^2]^{1/2}$, 
we find
 \begin{equation}
 H_{\bot}^{-1} ( x ) = \frac{H ( x )}{ 4 [ E^2 + | \tilde{\Delta} ( x ) |^2 ] }
 \; ,
 \label{eq:Hbotres}
 \end{equation}
i.e. $H_{\bot}^{-1} (  x)$ is proportional
to $H (  x)$.
To fix the constant $c_{n+1} ( x )$ in Eq.(\ref{eq:psinplus1}), we require
that the components of $ \sum_{i= 0}^{n+1}
\vec{\psi}_i ( x )$
satisfy the constraint (\ref{eq:constraint2}).
This implies
 \begin{equation}
 c_{n+1} ( x ) = - \frac{1}{2} \sum_{i=1}^{n} 
 \tilde{\psi}_{i}^{T} ( x ) 
 \vec{\psi}_{n+1-i} ( x )
 \; ,
 \label{eq:cnres}
 \end{equation}
where the vector $\tilde{\psi}_i (  x) $
is obtained from $\vec{\psi}_i ( x )$ 
by exchanging the first and third components and multiplying them
by $-1$, see Eq.(\ref{eq:psitildedef}). For odd $n$ we can show that
$c_{n} (x) = 0$.
We thus obtain an explicit  and very compact
recursive algorithm for
calculating the gradient expansion of the local DOS.
To zeroth order
the vector  $\vec{\psi} ( x )$ is given by Eq. (\ref{eq:psi02}). 
This corresponds to the {\it{adiabatic}} approximation
of elementary quantum mechanics.
The
step $n \rightarrow n+1$ is summarized as follows
 \begin{eqnarray}
 &\bullet& \ \vec{\psi}_{i} \text{ given for } i=0, \ldots, n \nonumber
 \\ 
 &\bullet& \ \vec{\psi}_{n+1} = - H_{\perp}^{-1} \partial_x
 \vec{\psi}_{n} - \frac{ \vec{\psi}_0 }{2}  \sum_{i=1}^{n}
 \tilde{\psi}_{i}^{T} \vec{\psi}_{n+1-i} 
 \; . 
 \label{eq:iteration}
 \end{eqnarray}
It is easy to implement this iterative algorithm  
on a symbolic manipulation program (such as {\em Mathematica}).
In this way the lowest few
terms in the gradient expansion can be obtained
in a straightforward manner.

Given the gradient expansion of the
local DOS (which can be directly obtained from the second
component of $\vec{\psi} ( x ) $), we can calculate the free energy
by simple integrations. 
Using the fact that in the normal state $R=1$,
the difference between the free energy densities
in the superconducting and normal state
of our effective one-dimensional  model 
at inverse temperature $\beta$ is given by $\nu_{1} f(x)$, where
 \begin{eqnarray}
 f (x) &=& -\frac{1}{\beta} \text{Re} \left[\int_{-\infty}^{\infty}
 d\omega \ \left[ R(x, \omega+i0) -1\right] \ln\left(1+e^{-\beta
 \omega}\right) \right] \nonumber \\
 &=& - \text{Re} \left[\int_{-\infty}^{\infty}
 d\omega \ \left[ N (x, \omega+i0) -\omega \right] \frac{1}{e^{\beta
 \omega}+1} \right] \nonumber \\
 &=& - \frac{2 \pi}{\beta} \text{Im} \left[
 \sum_{{\omega}_{n} > 0} N(x, i {\omega}_{n}) -i
 {\omega}_{n} \right]
 \; . 
 \label{eq:f1res}
 \end{eqnarray}
Here 
${\omega}_{n} = (2n+1) \pi /\beta$ are fermionic
Matsubara frequencies,
and 
 \begin{equation}
 N(x , \omega )=\int_{0}^{\omega} d \omega^{\prime}
 R(x , \omega^{\prime} )  
 \; .
 \end{equation}
Note that $\nu_{1} \text{Re} [N( x , \omega+i0)]$ is the integrated
one-dimensional local DOS. 
In Eq.(\ref{eq:f1res})
we have used the fact that
$R(x, \omega)$ is analytic in the upper half
of the complex $\omega$-plane.
The local DOS of the three-dimensional
superconductor is then given by
 \begin{equation}
 \rho ({\bf r} , \omega )  =  \nu_3
 \left< 
 {\rm Re} R_{\bf
  n}({\bf r} , \omega + i 0) \right>_{\bf n}
  \; ,
  \label{eq:rho3free}
  \end{equation}
where we have now written
$ R_{\bf
  n}({\bf r} , \omega )$ for $R ( x, \omega )$ to indicate
  the dependence on all parameters.
Similarly we write
$f_{\bf{n}} ( {\bf{r}} )$ for $f ( x )$ and
obtain for the difference between the free energies
of the three dimensional system in the superconducting
and normal state,
 \begin{eqnarray}
  F \{ \Delta ( {\bf{r}} ) \}  & =  &
  \int d^{3} {\bf{r}} 
  \left[ 
  \nu_3 
 \left< f_{\bf n}({\bf r}) \right>_{\bf n} 
  + \frac{|\Delta({\bf r})|^2}{\lambda} 
 \nonumber
 \right.
 \\
  &  &  \hspace{10mm} +
  \left.
  \frac{ [ {\bf{H}} ( {\bf{r}} ) - {\bf{H}}_e ( {\bf{r}} ) ]^2}{8 \pi}
  \right]
  \; .
 \end{eqnarray}
The second term is
the field energy of the superconducting pair potential (where
$\lambda$ is the coupling constant of the Gorkov electron-electron
interaction), and the last term is the magnetic field energy
of the superconductor
(where ${\bf{H}}_e ( {\bf{r}} )$
is the externally applied magnetic field\cite{Eilenberger64}).

The above equations
allow for a simple recursive calculation of the
gradient expansion of the free energy. 
To compare our results with Tewordt \cite{Tewordt64} and KKSL
\cite{Kosztin98,Kos98} we use
Eq.(\ref{eq:iteration}) to
calculate all terms in the gradient expansion up to fourth order.
Systematically adding total derivatives to the
expressions for the free energy (which do not
change the bulk properties of the superconductor),
we find the following expressions:

\vspace{7mm}

\widetext
\noindent
Zeroth order:
 \begin{equation}
 \rho^{(0)} ({\bf r}, \omega )  =  \nu_3 
 \Theta (\omega^2 - |\Delta|^2) 
 \frac{|\omega|}{\sqrt{\omega^2 - 
 |\Delta|^2}} 
 \; ,
 \end{equation}
 \begin{equation}
 F^{(0)} \{ \Delta ( {\bf{r}} ) \} 
  =   
 \int \ d^{3} {\bf{r }}
 \left[
 - \nu_3 
 \frac{2 \pi}{\beta} 
 \sum_{{\omega}_n > 0} 
 \left[
 \sqrt{{\omega}_n^2 + |\Delta|^2} - {\omega}_n \right]
 + \frac{|\Delta({\bf r})|^2}{\lambda} 
 + \frac{ [ {\bf{H}} ( {\bf{r}} ) - {\bf{H}}_e ( {\bf{r}} ) ]^2}{8 \pi}
 \right]
 \; .
 \end{equation}
Second order:
 \begin{equation}
 \rho^{(2)} ({\bf r} , \omega )  =  \nu_3 
 |\omega| \Theta (\omega^2 - |\Delta|^2) 
 \left<
 - \frac{5}{32} \frac{ 
 \left[ {\bf{n}} \cdot {\nabla}_{\bf{r}} 
 |\Delta|^2 \right]^2}{\left(\omega^2 - |\Delta|^2
 \right)^{\frac{7}{2}}} - \frac{1}{8} \frac{
 \left[ ( {\bf{n}} \cdot {\nabla}_{\bf r})^2 |\Delta|^2 \right] -
 3 | {\bf{n}} \cdot {\bf{{D}}}_{\bf{r}} 
 \Delta|^2}{\left(\omega^2 - |\Delta|^2 
 \right)^{\frac{5}{2}}}  \right>_{\bf n}    
 \; ,
 \label{eq:rho2}
 \end{equation}
 \begin{equation}
 F^{(2)} \{ \Delta ( {\bf{r}} ) \}
 =   \nu_3
 \int \ d^{3} {\bf{r}} 
 \frac{2 \pi}{\beta}  
 \sum_{{\omega}_n > 0} 
 \left<-\frac{1}{32}
   \frac{\left[ {\bf{n}} \cdot {\nabla}_{\bf r} 
   |\Delta|^2 \right]^2}{\left({\omega}_n^2 + |\Delta|^2
   \right)^{\frac{5}{2}}}
   + \frac{1}{8} \frac{| {\bf{n}} \cdot {\bf{D}}_{{\bf r}}
   \Delta|^2}{\left({\omega}_n^2 + |\Delta|^2 
   \right)^{\frac{3}{2}}} \right>_{\bf n}
   \; .
\end{equation}
\noindent
Fourth order:
{\footnotesize 
\begin{eqnarray}
& & \rho^{(4)} ({\bf r}, \omega )  = 
 \nu_3 
|\omega| \Theta (\omega^2 - |\Delta|^2) 
\left<\frac{1155}{2048} \frac{ \left[ {\bf{n}} \cdot {\nabla}_{\bf r}
       |\Delta|^2 \right]^4}{\left(\omega^2 - |\Delta|^2
     \right)^{\frac{13}{2}}}
     + \frac{42}{512} \frac{\left[ {\bf{n}} \cdot
             {\nabla}_{\bf r} |\Delta|^2 \right]^2 \left( 11 \left[ ( {\bf{n}} \cdot {\nabla}_{\bf
             r})^2 |\Delta|^2 \right]  - 15 | {\bf{n}} \cdot
         {\bf{D}}_{{\bf r}} 
     \Delta|^2 \right)}{\left(\omega^2 - |\Delta|^2
     \right)^{\frac{11}{2}}} \right. 
\nonumber \\
    & & \quad + \left. \frac{7}{128} \; \frac{ 5 
    | {\bf{n}} \cdot {\bf{D}}_{\bf r} \Delta|^4 
       + 4 \left[ {\bf{n}} \cdot
             {\nabla}_{\bf r} |\Delta|^2 \right] \left[ ({ \bf{n}} \cdot
             {\nabla}_{\bf r} )^3 |\Delta|^2 \right]
         -10 \left[{\bf{n}} \cdot {\nabla}_{\bf r} \left(| {\bf{n}}
             \cdot {\bf{D}}_{\bf r} \Delta|^2  
         \left[{\bf{n}} \cdot {\nabla}_{\bf
             r} |\Delta|^2 \right] \right) \right] + 3
             \left[ ( {\bf{n}} \cdot {\nabla}_{\bf
             r})^2 |\Delta|^2 \right]^2
       }{\left(\omega^2 - |\Delta|^2 \right)^{\frac{9}{2}}} \right.
\nonumber \\
    & & \quad + \left.\frac{1}{32} \frac{\left[ ( {\bf{n}} \cdot {\nabla}_{\bf
             r})^4 |\Delta|^2 \right] -5 \left[( {\bf{n}} \cdot
           {\nabla}_{\bf 
             r})^2 | {\bf{n}} \cdot
         {\bf{D}}_{{\bf r}} 
     \Delta|^2 \right]+5 | ( {\bf{n}} \cdot {\bf{D}}_{\bf r})^2
         \Delta|^2}{\left(\omega^2 - |\Delta|^2 \right)^{\frac{7}{2}}}
       \right>_{\bf n} 
       \; ,
       \label{eq:rho4}
 \end{eqnarray}
 \begin{eqnarray}
  F^{(4)} \left\{ \Delta ( {\bf{r}} ) \right\}
  & = & \nu_3 \int \ d^{3} {\bf{r}} 
  \frac{2 \pi}{\beta}  
  \sum_{{\omega}_n > 0} 
    \left<-\frac{35}{2048} 
    \frac{\left[ {\bf{n}} \cdot {\nabla}_{\bf r} 
       |\Delta|^2 \right]^4}{\left({\omega}_n^2 + |\Delta|^2
     \right)^{\frac{11}{2}}}
     + \frac{70}{512} 
     \frac{| {\bf{n}} \cdot {\bf{D}}_{{\bf r}} \Delta|^2 
     \left[ {\bf{n}} \cdot {\nabla}_{\bf r}
       |\Delta|^2\right]^2}{\left({\omega}_n^2 + |\Delta|^2
     \right)^{\frac{9}{2}}} \right. \nonumber \\
    & & + \left. \frac{1}{128}  \; 
    \frac{5 | {\bf{n}} \cdot {\bf{D}}_{\bf r} \Delta|^4
         -10 | {\bf{n}} \cdot {\bf{D}}_{\bf r} \Delta|^2 
         \left[ ( {\bf{n}} \cdot {\nabla}_{\bf
             r})^2  |\Delta|^2 \right] + 
             \left[ ({\bf{n}} \cdot {\nabla}_{\bf
             r})^2 |\Delta|^2 \right]^2
       }{\left({\omega}_n^2 + |\Delta|^2 \right)^{\frac{7}{2}}}   
       -\frac{1}{32} 
       \frac{| ({\bf{n}} \cdot {\bf{D}}_{\bf r})^2
         \Delta|^2}{\left({\omega}_n^2 + |\Delta|^2 \right)^{\frac{5}{2}}}
       \right>_{\bf n}  
       \label{eq:4thorder}
       \; .
       \label{eq:F4res}
\end{eqnarray}
}
\narrowtext
\noindent
Here $\Theta(x)$ is the step function, 
and ${\bf{D}}_{\bf{r}} =
 {\nabla}_{\bf{r}}
    + 2 i \frac{e}{c} {\bf A}({\bf r})$ is the covariant gradient. In
    deriving the above expressions we have used
\begin{equation}
 | ({\bf{n}} \cdot {\nabla}_{\bf r} )^{l} \tilde\Delta| = |({\bf{n}}
 \cdot {\bf{D}}_{\bf r})^{l} \Delta| \ , \quad l=0,1,2,\ldots \ .
\end{equation}
Note that all terms in our expansion are gauge invariant.
Since the terms involving
an odd number of derivatives cancel after directional averaging we
only presented 
the terms of even order. 
The directional averaging is easily done using
\begin{equation}
  \left< ({\bf n} \cdot {\bf{A}} )
  ( {\bf{n}} \cdot {\bf{B}} )
  \right>_{\bf n}  =  \frac{1}{3} {\bf A \cdot
    B} 
    \; ,
 \end{equation}
 \begin{eqnarray}
  \left< 
  ({\bf{n}} \cdot {\bf{A}} )
  ({\bf{n}} \cdot {\bf{B}} )
  ({\bf{n}} \cdot {\bf{C}} )
  ({\bf{n}} \cdot {\bf{D}} )
\right>_{\bf n} & = &
  \frac{1}{15} \left[ \left( {\bf A \cdot B} \right) \left( {\bf C \cdot
        D} \right) 
        \right.
\nonumber
\\
& & \hspace{-40mm} \left.
        +  \left( {\bf A \cdot C} \right)\left( {\bf B \cdot
        D} \right)
   + \left( {\bf A \cdot D} \right) \left( {\bf B
        \cdot C} \right) \right] 
 \; .
\end{eqnarray}
Because after directional averaging the above expressions
look even more complicated, we do not give the explicit results 
in this work.
Comparing Eq.(\ref{eq:F4res}) with the fourth order
correction to the free energy given by 
Tewordt\cite{Tewordt64} and by
KKSL\cite{Kosztin98,Kos98}, we find that our result
agrees with that of KKSL (up to a total derivative which
eliminates the first term on the right-hand-side 
of Eq. (\ref{eq:4thorder})).
We have not been able to find in the published
literature an explicit expression for the fourth
order correction to the gauge invariant local DOS, 
so that we cannot compare
Eq.(\ref{eq:rho4}) with the results of other authors.

\section{Conclusion} 
We have developed an efficient algorithm for calculating the gradient
expansion of the local DOS and the free energy of a
superconductor in an external magnetic field.
The linearization of the energy dispersion allows to
obtain the local DOS directly from the second component of a three
component wave-function ${\vec \psi (x)}$ satisfying a
pseudo-Schr\"odinger equation, which is formally identical to the
Schr\"odinger equation of a $J=1$ quantum spin in a time-dependent complex
magnetic 
field. The gradient-expansion of this
pseudo-Schr\"odinger equation turns out to be quite simple and
straightforward and can be easily implemented 
on a symbolic manipulation program such as {\em Mathematica}. It seems
to us that our algorithm gives an interesting alternative to the
algorithms developed by KKSL\cite{Kosztin98,Kos98}.
Our final result for the contribution to the gauge-invariant free energy
containing four gradients agrees with KKSL\cite{Kos98} and
is in disagreement with the expression given by
Tewordt\cite{Tewordt64}.
We believe that our algorithm could also prove advantageous in 
other problems where fluctuation effects can be mapped 
onto a $2 \times 2$ matrix equation for a Green's function
in external fields. One interesting example are two coupled 
Tomonaga-Luttinger models.

We would like to thank Simon Kos for pointing out to us that our
pseudo-Schr\"{o}dinger equation is in fact equivalent to the
Eilenberger equation.
This work was financially supported by the
DFG (Grants No. Ko 1442/3-1 and Ko 1442/4-1).

%

%
%
%
%
%
\end{document}